\documentclass{optica-article}

\journal{opticajournal} %

\articletype{Research Article}

\begin{document}

\title{Photon Number-Resolving Quantum Reservoir Computing}

\author{Sam Nerenberg,\authormark{1,\(\dag\)} Oliver D.\ Neill,\authormark{1,\(\dag\)} Giulia Marcucci,\authormark{1} and Daniele Faccio\authormark{1,*}}

\address{\authormark{1}School of Physics and Astronomy, University of Glasgow, G12 8QQ, United Kingdom.\\
\authormark{\(\dag\)}These authors contributed equally to this work.}

\email{\authormark{*}Daniele.Faccio@glasgow.ac.uk} %

\begin{abstract*} 
	Neuromorphic processors improve the efficiency of machine learning algorithms through the implementation of physical artificial neurons to perform computations. However, while efficient classical neuromorphic processors have been demonstrated in various forms, practical quantum neuromorphic platforms are still in the early stages of development. Here we propose a fixed, random optical network for photonic quantum machine learning in the specific form of a reservoir computer that is enabled by photon number-resolved detection of the output states. This provides access to a combinatorially-scaling Hilbert space, while using significantly simpler quantum states than comparable quantum machine learning approaches. The approach is implementable with currently available technology and lowers the entry barrier for quantum machine learning.
\end{abstract*}

\section{Introduction}\label{sec1}

Quantum machine learning~(QML) is a broad field in which the physical features of quantum systems are leveraged to process data~\cite{Biamonte2017}. In addition to promising speed-ups for classical tasks, QML systems are able to directly process quantum data leading to improvements in existing methods of extracting and manipulating quantum information~\cite{Ghosh2019,Preskill2023}. Quantum neural networks~(QNNs) are a subclass of QML in which artificial neural networks (ANNs) are formed from linked quantum states and their interactions, which may be parametrised and trained. Generally, these algorithms are implemented on quantum computers.
While various approaches to developing quantum computing hardware have been making great strides in recent years, the majority remain challenging to scale~\cite{Kjaergaard2020,Saida2024,Stassi2020,Acharya2024,Henriet2020,Evered2023,Moses2023}. The technology is out of reach for most laboratories and may be unnecessarily general for many routine data processing tasks which don't require guarantees of universality or strict fault tolerance.
This opens a door for neuromorphic hardware that can be specialised to various analysis tasks on classical or quantum data with minimal experimental overhead. 

Optical quantum computing platforms are one example which show promising scaling properties, due to the relative ease of building highly-multimodal optical systems, generating qubits and qudits through multi-photon states~\cite{Wang2019,Madsen2022,Arrazola2021,Bogaerts2020,Luo2023,Maring2024}.
Within optics, linear photonic networks~(LPNs) in particular have emerged as a promising candidate for implementing QNNs, with benefits including room temperature operation and low power consumption~\cite{McMahon2023}.
Importantly, despite their linearity, LPNs can generate complex nonlinear dynamics through suitable data encodings and decodings. In addition, LPNs have a rich theoretical framework in place which can be applied to a wide range of optical systems, and have already been used to demonstrate forms of quantum computation~\cite{Tan2019,Gard2014,Torma1995,vanderMeer,Steinbrecher2019,Wang2019,Broome2013,Madsen2022,MarcucciQuantum2020,Aaronson2010}.

Here we will consider the specific case of reservoir computing (RC), i.e.\ a particular subclass of ANN that has attracted attention as a viable and efficient neuromorphic platform. RC combines the collective hidden layers of an ANN into a single, random, high-dimensional layer with fixed dynamics and relies on the complexity of the feature map provided by this random layer to solve the task at hand. Training is achieved by performing a linear fit on the outputs of labelled data to learn a matrix of weights. The weight matrix is then multiplied with outputs generated by unseen data to make predictions.

Many physical systems are attractive candidates for RC due to their ability to generate rich, high-dimensional dynamics~\cite{TANAKA2019100}.
Optical systems are particularly promising for implementing RC, due to the density of optical modes in natural systems, their inherent parallelism, large frequency multiplexing potential, low latency, nearly dissipation-free dynamics and ability to passively perform large matrix operations~\cite{McMahon2023}.
Several systems have been developed based on random optical media whose properties are described by the theory of LPNs~\cite{Rafayelyan2020, GarciaBeni,Pierangeli2021,Biasi2023,Lupo2021}.
Such discrete, linear optical systems are ubiquitous in nature, while their unitary matrix representation makes them amenable to numerical simulation.
The scaling of the output space of an LPN implemented with classical light and detection is determined by the number of modes in the network, thus imposing a severe scaling constraint on classical approaches to {RC}. 
Applying quantum detection schemes can circumvent this limitation. In particular, using photon number resolving (PNR) detection, which allows the exact number of photons incident on a detector to be determined, in conjunction with photonic quantum states provides access to a Hilbert space which scales combinatorially in the number of modes and photons in the {LPN}.
There is, in principle, no limit to this scaling and increasing the number of photons allows the output space to grow rapidly without having to add extra spatial modes, mitigating the impact of losses present in larger networks.

Current methods of PNR can be categorised according to their method of operation: statistical analysis of multiplexed single-photon detection~\cite{You2020,Cheng2023,Fitch2003,Krishnaswamy2024,Achilles2004}, and low-noise systems able to distinguish signal features which scale discretely with photon number~\cite{Los2024,Hamamatsu2023,Schmidt2018}. In either case, high detector quantum efficiency is a key requirement for accurately measuring high photon-number states. These methods are early in their development but have recently progressed as far as commercial products, and detection efficiency and speed are ever-increasing with advances in detector technologies~\cite{Ceccarelli2021,Hummatov2023,Steinhauer2021,Hamamatsu2023}.

The overarching motivations to adopt quantum RC (QRC) are two-fold: the dimension of the output space of quantum systems typically scale rapidly with the number of constituent elements and that it can process quantum data directly. Other near-term approaches to quantum photonic RC have leveraged this. For example, by using novel optoelectronic components which exhibit memory effects at the single photon scale~\cite{Spagnolo2022} or by performing a random quantum walk in the space of orbital angular momentum states~\cite{Suprano2024}. However, in the first case experimental implementation of RC requires specialised integrated photonic circuits while in the second case, though demonstrated experimentally, the dimension of the output space can only grow linearly with the number of primitive optical elements~\cite{Liu2020}.

Here, we present an example of a general framework for photon number-resolving quantum reservoir computing (Photon-QuaRC). This is a QRC based on a linear photonic network in which information is encoded and manipulated in the input quantum states of light and read out with PNR detection. Our simulated model includes non-ideal aspects of realistic operation including detector losses and finite ensembles, which are necessary considerations for any practical experimental implementation. 
Furthermore, we show that we can harness the benefits gained through PNR without having to prepare large photon-number Fock states. The reservoir architecture avoids the need to optimise the network and reduces all training energy costs to a matrix pseudo-inversion. Our analysis considers general linear scattering processes and provides statistical evidence of robustness to different random reservoirs. This opens the door to implementation in a wide range of simple physical systems such as multimode fibres or scattering materials~\cite{Defienne2016,Rafayelyan2020}, thus providing a practical path towards versatile, scalable quantum machine learning.
\begin{figure*}[thb!]
    \begin{center}
        \includegraphics[width=1.0\columnwidth]{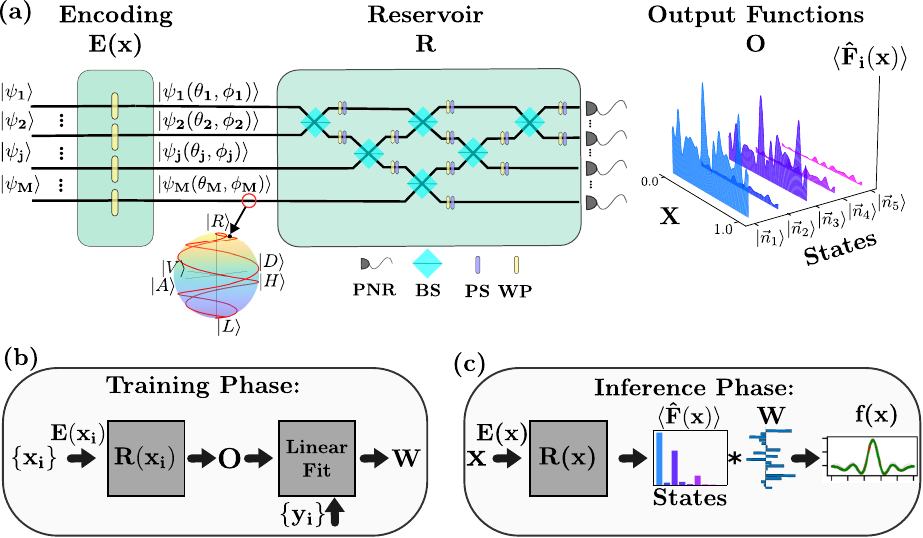}
        \caption{
            {\bf{Photon-QuaRC schematic overview.}}
            \textbf{(a)} Horizontally polarised quantum states of light \(|\psi_m \rangle\) in \(M\) different spatial modes are directed into an encoding block \(E(x)\)  that encodes the input data, \(x\), in a multimode polarisation state.
            The polarisation at each port is parametrised on the Poincar\'e sphere by  \(\theta_m(x)\) and \(\phi_m(x)\), the azimuth and zenith angles respectively. This state is fed into a random, fixed linear optical network built from variable beamsplitters (BS), waveplates (WP) and phase shifters (PS).
            Output states are subject to polarisation-independent measurement by PNR detectors. Approximation of the Fock state distribution \( \langle \hat{F}_i(x) \rangle\) by repeated sampling constitutes the device output, Eq.~(\ref{fxLPN}).
            \textbf{(b)}  During the training stage, labelled data \(\{x_i\}\) are encoded, and the reservoir outputs are stored in the design matrix \(O\). The elements of this array are then fit to labels \(\{y_i\}\) to learn weights, \(W\).
            \textbf{(c)} In the inference phase the estimated function \(f(x)\) is evaluated on new input data by multiplying \( \langle \Vec{\hat{F}}(x) \rangle\) by \(W\),  Eq.~(\ref{eq:linregr}).
        }\label{layoutFig}
    \end{center}
\end{figure*}

\section{Architecture}

RC is an umbrella term for a range of ANN systems which adopt a random and unoptimised hidden layer~\cite{TANAKA2019100,Lukoeviius2009,Butcher2013}. It is commonly divided into feed-forward (extreme learning machines, ELMs) and recurrent (echo state networks or liquid state machines) architectures~\cite{22_cucchi_}. The latter has some internal state which maintains memory of previous inputs either through feedback or through an adaptation of the hidden layer connectivity in response to inputs~\cite{Huang2006,Jaeger2010,Lukoeviius2009}. While recurrent architectures perform well on tasks which benefit from memory, such as processing time series data, there are many tasks where ELMs are sufficient. 
Additionally, existing ELMs often may be extended to process time series data or else adapted into echo state networks through simple structural changes~\cite{Rafayelyan2020,Butcher2013,Ortin2015}.
As a result, there has been growing interest in realising ELMs in quantum systems, launching investigations into the general properties and applications of such QML architectures~\cite{Innocenti2023,Xiong2025}.
We therefore do not discuss recurrence in this work and consider benchmark tasks which do not require memory.

Specifically, we consider a reservoir architecture where input information is encoded into the polarisation state of light, and a random network of waveplates and variable beamsplitters serves as the reservoir. It is important to note, however, that the Photon-QuaRC framework could be equivalently implemented with any suitable pairing of encoding scheme and linear optical reservoir. For example, encoding in phase with a multimode fibre reservoir, or dual-rail encoding and integrated photonic circuit reservoir. The benefits conferred by the use of quantum resources and PNR detection in the following treatment do not depend on the specifics of the physical system.

\subsection{Polarising linear optical networks as quantum reservoirs}
Figure~\ref{layoutFig}.a shows the schematic layout of our model for {Photon-QuaRC}.
A photonic quantum state \(|\psi\rangle\) is fed into the physical system which is composed of two sequential \(M\)-port LPNs: the encoding layer \(E(x)\) which encodes input data \(x\) in the polarisation degree of freedom of the input state, and the reservoir \(R\) which randomly couples both spatial and polarisation modes. 
These may be represented as unitary Fock space scattering matrices \(\mathcal{E}(x)\) and \(\mathcal{R}\) respectively,
that may be built from the mode-coupling matrices of the corresponding LPNs, as developed by Scheel~\cite{Scheel2004}.
The encoding does not couple spatial modes, acting only on the polarisation degree of freedom to generate trajectories on a Poincar\'e sphere, schematically indicated in the figure.

The output of the QRC after PNR detection can be written as
\begin{equation} 
    \langle \hat{F}_i(x) \rangle =\langle \psi\mathcal{E^{\dag}}(x)\mathcal{R^{\dag}}|\hat{F}_i |\mathcal{RE}(x)\psi \rangle,
\label{fxLPN}
\end{equation}
a probability mass function over Fock state observables \({\hat{F}}_i(x) = |\vec{n}_i\rangle \langle \vec{n}_i|\), where the \(\vec{n}_i\) denote all \(d = \binom{N+M-1}{N}\) possible occupations of \(M\) modes by \(N\) photons.
The scaling of \(d\) with \(N\) is key as it allows us to expand the feature space combinatorially faster than the physical network complexity.

Prior LPN-based approaches to QML used networks made of trainable blocks that are tuned to produce a desired target function~\cite{Gan2022}. In contrast, a key feature of the Photon-QuaRC architecture is that the target function is computed without having to optimise over the network parameters.
Instead, we perform a classical linear regression over the QRC outputs from labelled training data to produce a weight matrix  \(W\). These weights provide an approximation \(f(x)\) to a target function \(f^*(x)\) using the reservoir output through a single matrix multiplication,
\begin{equation}
    f(x) = W\langle \Vec{\hat{F}}(x) \rangle,
\label{eq:linregr}
\end{equation}
where \(W\) is the weight matrix learned from linear regression on labelled data, shown in Figure~\ref{layoutFig}.b.
Figure~\ref{layoutFig}.c illustrates inference with the learned weights on unseen data.

This approach is similar to that of quantum kernel methods (QKM) where a fixed (but not random) network is used to encode data points into photonic quantum states and then physically evaluate their pair-wise inner products~\cite{Yin2024,Bartkiewicz2020}. These quantum circuits provide an implicit nonlinear projection by calculating the relative distances between data points in a high-dimensional Hilbert space. Once the so-called kernel matrix of inner products is determined, a classical classifier such as a support vector machine is trained to make predictions on unseen data. In contrast, Photon-QuaRC calculates the projection into the Hilbert space directly. QKMs have the benefit of rigorous theoretical performance guarantees~\cite{Liu2021}, whilst Photon-QuaRC has potential advantages in the near-term. For each new evaluation data point, QKM requires the inner product with every training point, yielding an inference complexity that scales with the size of the training set. Photon-QuaRC distils the training information into the \(W\) matrix, giving constant inference complexity. In addition, photonic circuits for calculating inner products can be difficult to scale to larger photon numbers, whereas direct detection of multiphoton output states can be readily accomplished with detector arrays.

One of the key features of both approaches is that they move the optimisation to a fully classical and convex post-processing step ensuring efficient trainability. Methods for calculating gradients of parametrised quantum models have been developed making use of parameter-shift rules~\cite{Wierichs2021}. However, at present, attempts to train such models via gradient methods such as back-propagation have failed to scale except in cases of extremely specific architectures~\cite{Bowles2023}. In addition, optimisation of these models is, in general, non-convex and efforts to scale to larger problem sizes are plagued by the issue of barren plateaus in which the loss landscape becomes exponentially flat~\cite{Larocca2024}. The classical optimisation methods in Photon-QuaRC and QKM side-step these problems and ensure trainability making them compelling alternatives for QML devices in the near term.

\subsection{Polarising network model}

We construct the network \(R\) adapting the method proposed by Reck et al.~\cite{Reck1994} with some key modifications: we replace each spatial mode with a pair of corresponding polarisation modes and the constituent optical elements are variable beamsplitters, arbitrary birefringent waveplates and phase shifters. The full network then couples 2M modes, and the primitive elements operate on \(4\times4\) subspaces (see the SM for details).
 
The input to the QRC is a fixed state which is subsequently encoded with the data and scattered into a superposition of Fock states at the output. To describe the action of the encoding layer at each input port we apply the ladder operator replacement rule
\begin{equation}
\hat{a}_m^{\dag} \rightarrow \cos{\theta_m(x)}\hat{a}_{m,H}^{\dag} +\sin{\theta_m(x)}e^{i\phi_m(x)}\hat{a}_{m,V}^{\dag},
    \label{PolState}
\end{equation}
where \(a_{m,H/V}^{\dag}\) denotes a creation operator acting on the \(H/V\) polarised component of port \(m\) and \(\theta_m\) and \(\phi_m\) are the mixing and phase angles, respectively. Thus, each data point is encoded as a set of Stokes vectors by waveplates at each input port. A given encoding scheme consists of a joint trajectory on the Poincar\'e sphere parametrised by the data. As the input state traverses this trajectory on the sphere, the probability of each output state follows a smooth curve as a result of Eq.~(\ref{PolState}). These output functions are the basis elements which are combined in Eq.~(\ref{eq:linregr}) to perform a given computational task, e.g.\ the approximation of a target function. 

\subsection{Reservoir characterisation}
Previous work has shown that the output of many quantum neural networks, particularly those whose structure consists of unitary operations parametrised by classical data such as ours, can be expressed as a partial Fourier series of the data over a restricted set of frequencies~\cite{Gan2022,Schuld2021,GilVidal2020,Wierichs2021,Xiong2025}.
This output \(g(x)\) can be written as
\begin{equation}
    g(x) = \sum_{\omega \in \Omega_n} c_\omega e^{i\omega x},
    \label{fseries}
\end{equation}
where \(\Omega_n\) is the set of all unique frequencies that the network can generate. Frequencies \(\omega\in\Omega_n\) are determined by the encoding scheme, while the coefficients \(c_\omega\) are fixed by the network structure. In the special case of photonic QNNs, frequencies are also determined by the number of input photons~\cite{Gan2022}. The properties of \(\Omega_n\) (cardinality, bandwidth) in part determine the class of functions that the network can approximate, a.k.a.\ \textit{expressivity}. Note that while \(\Omega_n\) contains information about the ability of a given design to specialise to a task, it does not fully characterise the performance nor robustness to realistic conditions of operation.

This task specific performance depends on the span of the observable functions the reservoir can generate and whether they can be feasibly detected above noise. We therefore introduce additional measures beyond the Fourier spectrum to better quantify expressivity. During the training phase, labelled data is fed into the network and the reservoir output is stored in an array \(O_{ij} = \langle \hat{F}_i(x_j) \rangle\). The Gram matrix \(OO^{\dag}\) then contains all inner products between output function pairs and its rank determines the number of unique basis functions that span the output space. Although the output functions contained in \(O\) can be mapped to the elements of Eq.~(\ref{fseries}) using a Fourier transform, the advantage of using \(\text{rank}(OO^{\dag})\) as an expressivity metric is that it counts the dimension of the space spanned by the observables rather than the number of frequencies in the output. 
This also implies that in order to be fully expressive, a network must output at least as many observables as the number of frequency components it can generate.

In a realistic scenario \( \langle \hat{F_i}(x) \rangle\) must be approximated by injecting a finite number of copies of the input state, and the output will be collected with imperfect detectors. As it is a non-uniform distribution, we will only be able to access the output features which occur with the highest probabilities. Some basis functions will not be distinguishable above noise with a practical number of samples, reducing the number of useful output features. To establish an effective rank, we take the singular value decomposition of \(OO^{\dag}\) to find  the normalised singular value spectrum of the matrix, \(\sigma_i\). We then count the number of singular values with magnitude above a threshold equal to \(k\) times the reciprocal of the total number of samples. This number gives us the conditioned rank \(R_c\) which is the number of linearly independent output features which are sampled on average, at least \(k\) times.

\subsection{Encoding schemes}
The expressive power of a network is determined by the encoding scheme and the input state. While an exhaustive investigation of all possible encodings and input states is outside the scope of this work, we will discuss a few schemes here, also shown in the top row of Figure~\ref{OutputsvsEncodings} (see the SM for more details).

{\emph{Uniform linear:}}
The simplest encoding is one in which all photons are given the same linear polarisation angle \(\theta \in [0,2\pi]\) which traverses one equatorial orbit over the data domain, \(\theta = lx\) for scaling factor \(l\).

{\emph{Multi-slope linear:}}
The uniform linear scheme is ultimately limited by the number of photons, so it is natural to try to broaden the spectrum by encoding multiple frequencies directly in each port. One method is to apply a different linear function to the polarisation of each spatial mode \(\theta_m = l_m x\).

{\emph{Spiral:}}
By including elliptical polarisations such that our data are encoded in both \(\theta\) and \(\phi\) in Eq.~(\ref{PolState}), we can design encodings with useful properties inherited from the topology of the Poincar\'e sphere. A natural geometry is a spiral which traverses the sphere from pole to pole and returns to form a closed curve. This encoding can undergo an arbitrary number of azimuthal revolutions \(l_m\) over the domain of the data while also providing low-frequency content from the zenith traversal.

Beyond the specific method for encoding information on to the photon states, we also consider the nature of the photon states themselves. Optical QML systems typically require generating large-photon-number Fock states. This leads us to consider how high-photon-numbers might be achieved using simpler quantum resources than Fock states and whether they yield comparable performance.
For example, prior work has shown that through combinations of single-photon Fock and coherent states, one can create quantum mechanical states with arbitrary photon numbers~\cite{2010_silberberg,Windhager2010,Barbieri2010,Lvovsky2002,Zavatta2004}. This approach also requires photon number-resolving detection and post-selection to distil quantum interference of the desired order from the output. 

We adopt this concept and start from a general state that combines coherent and Fock components in a unified form,
\begin{equation}
  |\psi_{\vec{\alpha},\vec{n}} \rangle = 
  \prod_{m=1}^M \frac{{(\hat{a}_m^{\dag})}^{n_m}e^{\alpha_m \hat{a}_m^{\dag} - \alpha_m^* \hat{a}_m}}{\sqrt{L_{n_m}(-{|\alpha|}^2)n_m!}}|0 \rangle,
\label{hybridstate}
\end{equation}
where \(L_n(x)\) is the Laguerre polynomial of degree \(n\), \(\hat{a}_m^{\dag}\) is a creation operator acting on port \(m\) and \(\alpha_m\) and \(n_m\) are the coherent state amplitude and number of added photons in port \(m\) respectively. If the \(\alpha_m\) or \(n_m\) are set to zero, Eq.~(\ref{hybridstate}) simplifies to the standard form of a Fock state or a coherent state, respectively. Any state with non-zero \(n_m\) and \(\alpha_m\) is referred to as a hybrid state. In what follows we will notate a hybrid state as \(|\psi_{\vec{\alpha},\vec{n}} \rangle=|\alpha_1 + n_1,\ldots,\alpha_m + n_m \rangle\) (see SM for details). 
We also consider fully distinguishable Fock states, meaning the constituent photon wave functions do not overlap~\cite{Tichy2013}. Experimentally, this distinguishability may arise from factors such as time delays or wavelength separation. Distinguishable states therefore do not undergo multiphoton interference in the reservoir.
We will denote a distinguishable Fock state as \(|\vec{n}_d \rangle=|n_{1,d},\ldots,n_{m,d} \rangle\).

\subsection{Simulation details}
In order to simulate the reservoir computer, we use the implementation of the `strong linear optical simulator' algorithm from the Perceval Python library~\cite{heurtel2023perceval, 23_heurtel_slos}. We construct reservoirs by uniformly sampling wave-plate angles and beamsplitter parameters at each mode-coupling point in the reservoir, in accordance with the beamsplitter validity conditions~\cite{Uppu:16}. All the couplings are combined to form the scattering matrix in the Scheel formalism which is used to compute scattering probabilities for given input states~\cite{Scheel2004}. 
We consider the reservoir itself to be lossless and post-select on observed number states, tracing out the polarisation degree of freedom. We do however consider detection dark-counts and quantum efficiency. Further discussion and details of the analytical model may be found in the Supplementary Materials.

\section{Results}
\subsection{Feature-space scaling}

\begin{figure}[t!]
  \begin{center}
  \includegraphics[width=0.9\columnwidth]{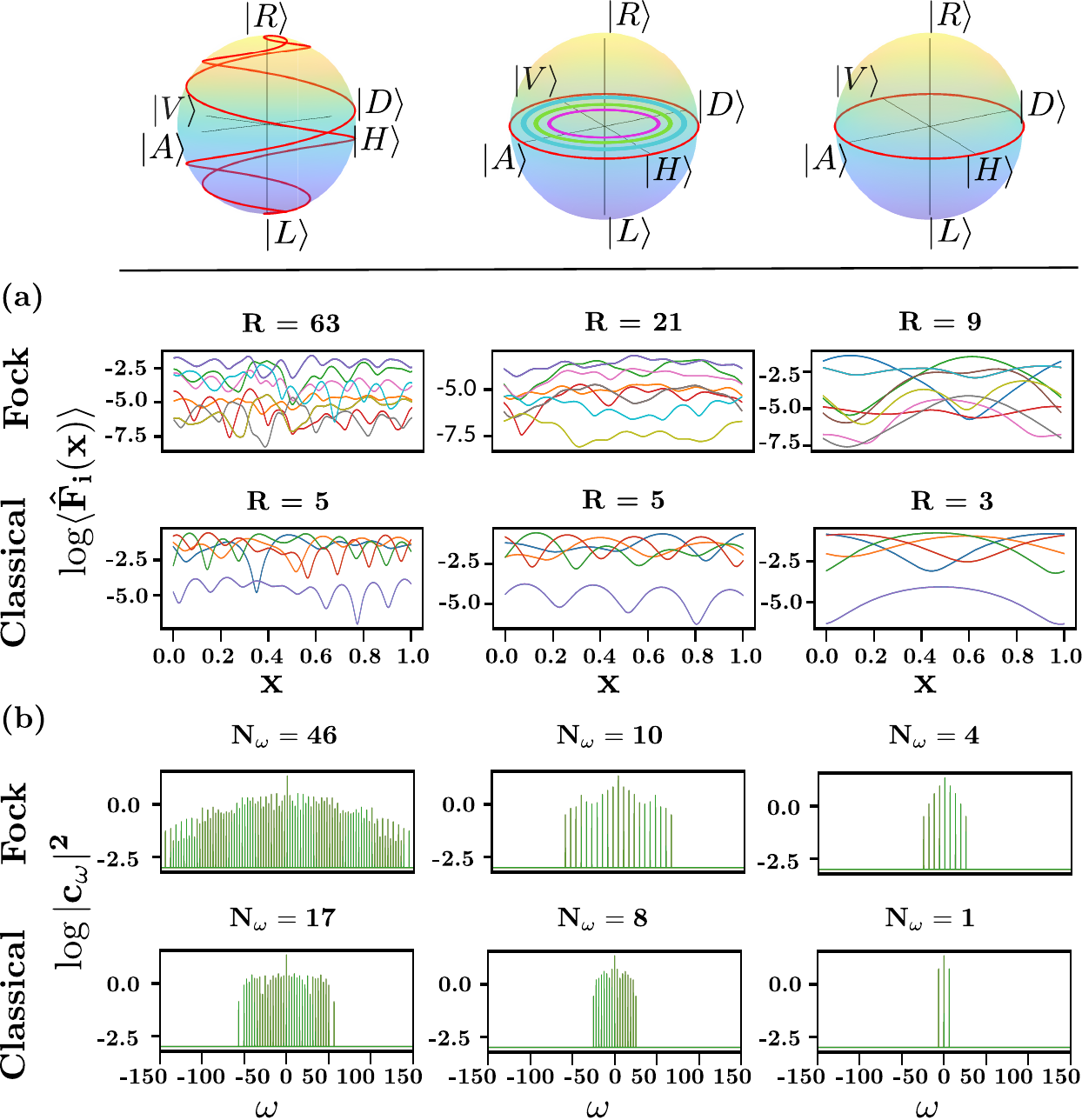}
  \caption{{\bf{Photon-QuaRC, quantum and classical behaviour for various encodings.}} Two different states and detection schemes are applied to the same \(M=5\) network. {\bf Fock PNR:} A four-photon Fock state \(|\vec{n}\rangle = |1,1,1,1,0 \rangle\) with perfect PNR detection. {\bf Coherent Intensity:} A four-port classical state \(|\vec{\alpha}\rangle = |.5,.5,.5,.5,0 \rangle\) with intensity detection at each port.
  \textbf{(a)} The first 10 most probable output functions for each case.
  \textbf{(b)} Corresponding Fourier spectra for each set of output functions. Each column corresponds to a different encoding scheme. {\bf Left:} Spiral with \(l_m = 4m\). {\bf Center:} Multi-linear where \(l_m = m\). {\bf Right:} Uniform-linear. The label \(R\) refers to the rank of the distribution.
  \(N_{\omega}\) denotes the number of elements in a spectrum excluding redundant negative frequencies and the DC component.}\label{OutputsvsEncodings}
  \end{center}
\end{figure}

Figure~\ref{OutputsvsEncodings} compares the characteristics of the output functions and Fourier spectra generated by the same \(M = 5\) mode reservoir under fully quantum or fully classical operation. 

We show the output functions (Fig.~\ref{OutputsvsEncodings}.a) and frequency spectrum (Fig.~\ref{OutputsvsEncodings}.b) generated by the different encoding schemes for both a four photon Fock state with PNR detection and a four port coherent state with intensity detection (see SM for details of choice of coherent states' \(\alpha\)). The choice of encoding can tune the rank (indicated as \(\mathbf{R}\) in the figures), distribution, and spectrum of the output. In the Fock with PNR case, adding complexity to the encoding trajectory increases the size of the Fourier spectrum as well as the rank of the output. This in turn improves the expressivity of the system. In contrast, the rank of the classical case - coherent states with intensity detection - is limited to the number of output ports of the network and the Fourier spectrum growth is much slower.
Thus, by using PNR detection we gain an {improvement in expressivity} over fully classical operation through the ability to scale both the number of output functions and Fourier components, increasing the likelihood of generating a larger set of independent basis elements.

\subsection{Photon-QuaRC as a function interpolator}\label{sec:interp}
\begin{figure*}[t!]
\begin{center}
\includegraphics[width=1.0\columnwidth]{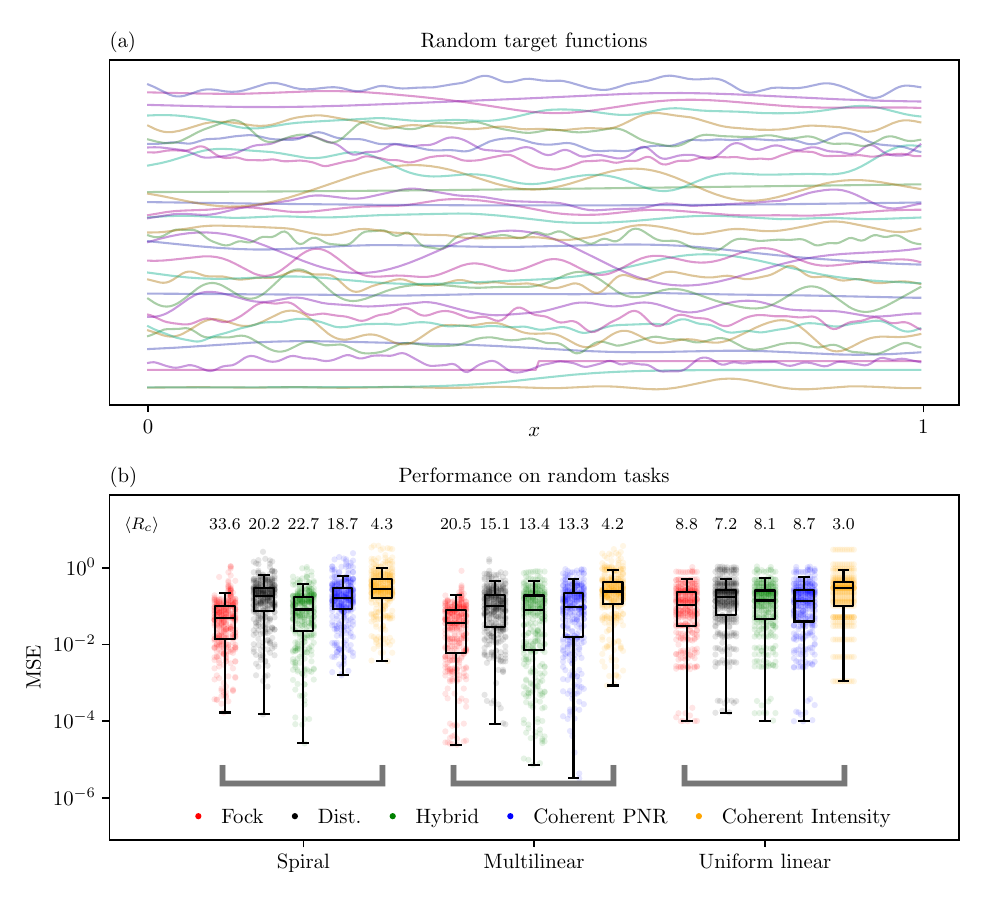}
\caption{{\bf{Random function interpolation statistics.}} (a) Set of 35 randomly generated target functions. Each function is generated by uniformly sampling amplitudes, phases and frequencies within a fixed bandwidth. DC offsets are added here to improve plot visibility. (b) Box plots of mean squared error for prediction on target functions, for each combination of input state and encoding, with 10 random reservoir instantiations. Average conditioned ranks (over reservoirs) \(\langle R_c \rangle\) are listed  for each case. The encodings are {\bf Spiral} with \(l_m = 4m\). {\bf Multi-linear} where \(l_m = m\). {\bf Uniform-linear} where \(l_m = 1\). The states used are {\bf Fock:} \(|\vec{n} \rangle=|1,1,1,1,0 \rangle\), \textbf{Distinguishable:} \(|\vec{n}_d\rangle = |1_d,1_d,1_d,1_d,0 \rangle\), {\bf Hybrid:} \(|\psi_{\vec{\alpha},\vec{n}} \rangle=|0.5 + 1,0.5 + 1,0,0,0 \rangle\), {\bf Coherent PNR:} \(|\vec{\alpha} \rangle = |0.5,0.5,0,0,0\rangle\), all with PNR detection, and {\bf Coherent Intensity:} Fully classical operation using \(|\vec{\alpha} \rangle = |0.5,0.5,0.5,0.5,0\rangle\) with intensity detection. Each coherent state subject to PNR detection is approximated as a superposition of Fock states up to the sixth order. All trials use post-selection on \(\leq 4\) photon events with \(\eta = .9\), \(N_{samp} = 10^7\) and a 50/50 test/train data split.}\label{rand_interp}
\end{center}
\end{figure*}

In Figure~\ref{rand_interp}, we demonstrate differences in performance between various implementations applied to function interpolation machine learning tasks. 

We first define the target function \(f^*(x)\) on an interval \(x \in [0,1]\), and discretise it to generate sets of data and labels, \(\{ x_i, y_i \}\). These are randomly and uniformly split into two sets for training and validation, such that the system must learn the entire function in order to correctly interpolate it.
To simulate the output in real-world scenarios, we need to account for imperfect detection and limited sampling.
A detector model with quantum efficiency \(\eta\) is first applied to the ideal distribution \(\langle \hat{F_i}(x) \rangle\) (See Supplementary Section 8).
We then post-select the Fock subspace that corresponds to the detection events of interest while all other events are grouped into a single reject state. For each data point \(x\), we fix the number of copies of the input state \(N_{samp}\) that we use to sample the output distribution including the reject state.
Once this number of samples has been drawn, we remove the reject state from the distribution and renormalise to obtain our approximation to \(\langle \hat{F_i}(x) \rangle\).\@ A discussion of how post-selection is used as a hyperparameter in Photon-QuaRC may be found in Supplementary Section 9.
The training data are used to form the design matrix \(O\) from which we extract the rank and derive weights \(W\). 
We then evaluate the fit on the validation data and evaluate a mean squared error (MSE) between the outputs and their corresponding labels to quantify the task performance.

\subsection{Performance enhancement with quantum resources}
\begin{figure*}[!hbt]
\begin{center}
\includegraphics[width=1.0\columnwidth]{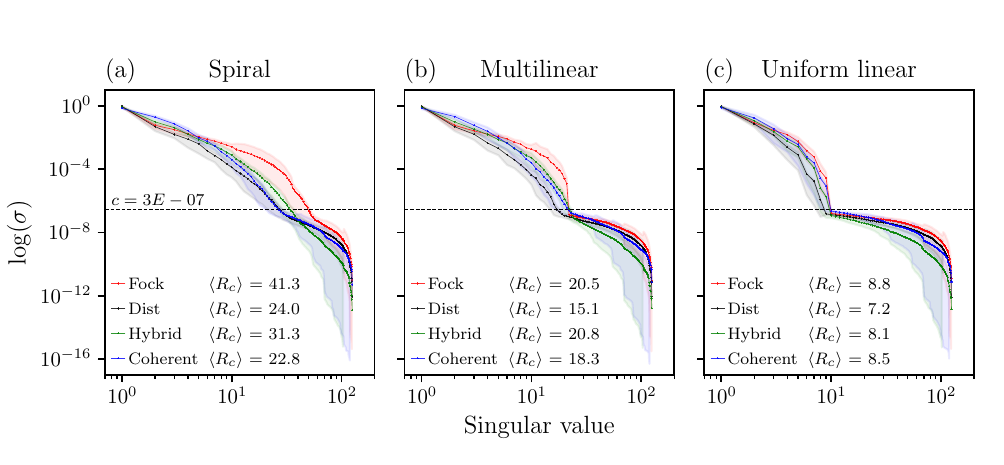}
\caption{\textbf{Singular value spectra} for a combination of states and encodings over ten different random reservoir realisations. The spectra are normalised to sum to one. The bold line indicates the average over the reservoirs for each case while the shaded regions show the deviation. The average conditioned rank \(\langle R_c \rangle\) is calculated from the number of singular values which fall above the cut-off threshold \(c\) shown here as the horizontal dashed line. All spectra are calculated with parameters \(\eta = .9\),  \(N_\text{samp} = 10^7\) and post-selected for detection events of \(\leq 4\) photons. The states used are \textbf{Fock:} \(|\vec{n}\rangle = |1,1,1,1,0 \rangle\), \textbf{Distinguishable:} \(|\vec{n}_d\rangle = |1_d,1_d,1_d,1_d,0 \rangle\), \textbf{Hybrid:} \(|\psi_{\vec{\alpha},\vec{n}} \rangle = |.5 + 1, .5 + 1,0,0,0 \rangle\) and \textbf{Coherent:} \(|\vec{\alpha}\rangle = | 1.5,1.5,0,0,0 \rangle\). Encodings are (a) \textbf{Spiral} with \(l_m = 4m\) (b) \textbf{Multi-Linear} with \(l_m = m\) and (c) \textbf{Uniform linear} with \(l_m = 1\).}\label{svalspect}
\end{center}
\end{figure*}

The ability of a given design to fit a specific function ultimately depends on the overlap of the output and target spectra. As such, we analyse the statistics of MSE over random tasks displayed in Fig.~\ref{rand_interp}.a to pick out the primary determinants of performance in a task-independent manner. We can see that at the low extremes, conditioned rank predicts poor performance, but once the rank exceeds a sufficient dimension it no longer predicts relative performance. Due to the fixed size of the network, the fully classical case (coherent intensity) can never reach this threshold dimension and consistently performs the worst. This demonstrates that PNR is the primary quantum resource that determines performance and, without it, improvements due to other factors are minimal. We see that the differentiation between states increases dramatically as we move from uniform linear to more complex encodings. This implies that the encoding must supply enough complexity such that the different physics of the input states can begin to show their {differences in performance}. Although average performance is upper and lower bounded by Fock and distinguishable states respectively among the PNR cases, there is no absolute hierarchy of performance with increasing quantum resources at the input. 
The only difference between the distinguishable and Fock states is the presence of multiphoton interference, which contributes a greater diversity of output frequencies in the latter case, demonstrating the benefit of quantum resources beyond {PNR}.
To see detailed fitting results for typical functions used to benchmark RCs, consult Supplementary Figure S4.
 
Representing the output functions in the basis of the singular vectors of \(OO^{\dag}\) allows us to separate linearly independent features and their relative probabilities. Figure~\ref{svalspect} displays the singular values, normalised to sum to one, for all combinations of the four input states with PNR and three encodings. Notably, these spectra have a characteristic shape consistent with random complex networks which allows us to analyse some of their features generally rather than on a case-by-case basis~\cite{Thibeault2024}. The noise level, determined by \(N_{samp}\) can be seen in Figure~\ref{svalspect} as the inflection point in each curve near an ordinate value of \(3 \times 10^{-7}\). The abscissa value at which each curve intersects this threshold establishes a conditioned rank, beyond which singular vectors are poorly sampled and contribute mostly noise. The spectra of each input state display markedly different behaviour. The Fock states have a flatter distribution and fall off the slowest despite having lower initial values than the coherent or hybrid states. The distinguishable states notably fall the most rapidly while the coherent and hybrid trade places depending on the encoding. A key observation is that, while \(R_c\) is not robustly predictive of performance, by inspection we can relate performance to the spectra where, in each case, a flatter spectrum yields a lower average {MSE}.
Flatter spectra correspond to reservoirs that generate more independent basis functions above the sampling noise floor. 
Quantifying this feature is challenging due to the log scale, as metrics of uniformity such as Shannon entropy or Gini coefficient are heavily biased by the large initial values. 
It is likely that multiple metrics must be combined to capture a robust task-independent performance hierarchy, and research into this is ongoing. 

Though Fock states yield the best performance in our tasks, one can obtain a given number of samples much faster from coherent sources than Fock sources of the same average photon number using the current experimental state-of-the art~\cite{Maring2024}. Thus, despite being less efficient in terms of total number of experimental repetitions required, a source where state copies are cheap (e.g.\ a coherent state) may be more appropriate for near-term applications. Aside from the potential increase in performance with quantum resources at the input, we emphasise that in all cases we see a significant increase in performance through photon number-resolving detection at the output: this alone grants {a performance improvement} over the fully classical case - i.e.\ coherent states without PNR detection.

\section{Image classification}

In order to demonstrate the performance of Photon-QuaRC on a more complex task, we apply the system to image classification. We use the standard digit MNIST dataset~\cite{10_lecun_mnist} and a set of six datasets from the MedMNIST ensemble~\cite{medmnistv2}. The full details of the classification scheme can be found in the Supplementary Section 5.

Figure~\ref{fig:image_accuracy} shows the classification accuracy on both train and test sets for each dataset, for the same Fock, coherent PNR and coherent intensity states used in Section~\ref{sec:interp}. We also show the ``majority'' classifier, which always predicts the most likely class from the training dataset, and represents the lower bound in performance, allowing comparison of unbalanced datasets. In a balanced dataset of \(N\) classes, we expect the majority accuracy to be \(1/N\).

\begin{figure*}[thb!]
	\begin{center}
			\includegraphics[width=1.0\columnwidth, trim={0 6.35cm 0 0}, clip]{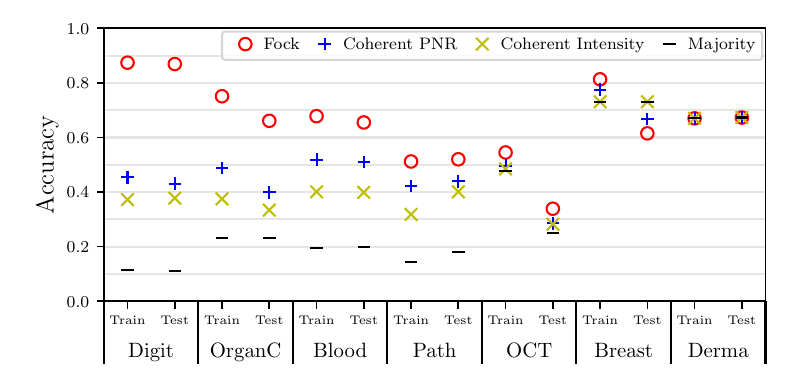}
			\includegraphics[width=1.0\columnwidth]{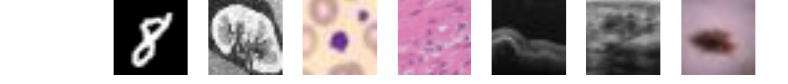}
			\caption{
					{\bf{Image classification accuracy.}} Classification accuracy on the train and test sets for digit MNIST and a subset of the MedMNIST datasets. The states used are Fock, coherent PNR and coherent intensity (see Section~\ref{sec:interp}). The majority classifier is shown as a lower bound.
          Example images from each dataset given below.
			}\label{fig:image_accuracy}
	\end{center}
\end{figure*}

We note that the encoding strategy used here, and detailed in the SM is generic i.e., has not been optimised specifically for these tasks.
This lack of optimisation can  be seen for example in three datasets (OCT-, Breast- and Derma-MNIST) which exhibit reduced performance with this particular encoding scheme and reservoir configuration.
These datasets are harder to classify correctly, in part due to heavily imbalanced classes, which is reflected in the performance of the majority classifier. In the case of BreastMNIST, we also see that the classifier overfits to the training data, resulting in worse performance on the test set than the majority classifier.
The Matthews correlation coefficient (MCC) allows us to compensate for imbalanced datasets, and is shown in Supplementary Figure S6.
Despite these edge cases and encoding limitations, there remains a clear ordering where states subject to PNR perform better than the fully classical case, and Fock states outperform coherent states---matching the results found in the interpolation tasks.

\section{Conclusions}

We have presented an approach to QML that is based on a random linear photonic network. This shows PNR detection as the minimal quantum resource alone yields significant improvement over classical (intensity only) detection, and without further classical post-processing. Beyond the use of PNR, we found that increasing quantum resources at the input can confer an additional performance improvement through multiphoton interference.

Although this performance improvement {hints at} a so-called quantum advantage relative to purely classical operation, recent works have suggested that this might not manifest as the system scales. Despite the scaling potential of quantum ELMs and QKMs, there is an exponential trend of the output to concentrate to a narrow range of outputs independent of the input data as the system size grows~\cite{Xiong2025,Thanasilp2024}. It is not clear whether this would limit Photon-QuaRC in practice. From a theoretical standpoint, the analysis in~\cite{Xiong2025} assumes that the number of output observables scales polynomially while in Photon-QuaRC the number of possible output observables, in principle, scales directly with the Hilbert space: combinatorially. As addressed by Xiong et al., combining this scaling with recent developments in measurement protocols such as classical shadows may allow Photon-QuaRC to operate with polynomially many samples~\cite{Elben2023,Huang2021,Innocenti2023b}. This is a promising direction for future research. However, from a practical perspective many tasks of interest may not require a Hilbert space size at which concentration hinders their evaluation. In the analysis presented here, tasks of a useful level of complexity were demonstrated using a modest number of photons and spatial modes. For any task, one should only use a Hilbert space large enough to contain the necessary features.

Furthermore, neuromorphics and specifically ELMs, are inherently task specific devices and a given architecture should be used for the purposes to which it is naturally suited. Photon-QuaRC natively produces finite Fourier series which can be scaled in bandwidth and tuned to the task of interest via encoding, post-selection and input photon number. Despite being especially suitable for representing nonlinear band limited functions, we have shown that our system is also applicable to  high dimensional nonlinear tasks such as image classification. We expect that the method will also be well suited to tasks in which data encoded in quantum states are mapped to functions of few variables, such as integrated quantum information processing units in communication networks. 

Photon-QuaRC is distinguished by a set of practical features that make it experimentally accessible. The reservoir architecture allows the device to be realised with simple optical systems such as multimode fibres, beamsplitter arrays, or integrated photonic circuits, and saves time and energy in the training stage.%
While we present results from a simulated polarising beamsplitter network, the generality of our analysis and the ubiquity of LPNs leaves many possible realisations for future work. It is PNR detection which is the key resource that provides a {performance improvement compared to non-PNR schemes}.
The ability to maintain good performance without generating high photon number Fock states, along with developments in PNR detection technologies, make this QML approach comparatively accessible and therefore an enabling feature for future generation quantum neuromorphic systems.

\section*{Funding}
The authors acknowledge financial support from the Royal Academy of Engineering Chairs in Emerging Technologies and the UK Engineering and Physical Sciences Research Council (projects no. EP/T00097X/1, EP/Y029097/1).

\section*{Acknowledgements}
The authors would like to thank Ilya Starshynov for insightful discussions.
Author contributions: S. Nerenberg: conceptualisation, methodology, investigation, formal analysis, visualisation, writing -- original draft. O. Neill: methodology, investigation, formal analysis, visualisation, writing -- original draft. G. Marcucci: methodology, writing -- review and editing. D. Faccio: conceptualisation, supervision, funding acquisition, writing -- review and editing.

\section*{Disclosures}
The authors declare no conflicts of interest.

\section*{Data availability}
Code replicating all results in this paper is available at \href{https://github.com/odneill/qrc}{https://github.com/odneill/qrc}, archived at \href{https://doi.org/10.5281/zenodo.15113065}{doi:10.5281/zenodo.15113065}~\cite{neill25_qrc}. 

\section*{Supplemental document}
See Supplementary Materials for supporting content.

\end{document}

\maketitle

\section{Representation of number states of polarised photons}\label{sec:fock_states}
We can write a state with definite photon number for a polarised device with M ports and N photons - each with polarisation state \((\theta_{m,k},\phi_{m,k})\) in Poincar\'e sphere coordinates - as
\begin{equation}
    |\Psi_{\Vec{n}}(\boldsymbol {\theta},\boldsymbol {\phi})\rangle = 
    \prod_{m=1}^{M}\frac{\prod_{k=1}^{n_m}a_m^{\dag}(\theta_{m,k},\phi_{m,k})}{\sqrt{n_m!}}|0 \rangle,
    \label{MportRepDiffPol}
\end{equation}
where \(a_m^{\dag}(\theta_{m,k},\phi_{m,k})\) is a creation operator acting on port \(m\) imparting an internal state \(k\), and \(n_m\) is the number of photons in port \(m\) such that \(\sum_{m=1}^M n_m = N\). The outer product runs over the ports and the inner applies the appropriate creation operators at that port. For simplicity as well as experimental practicality we will assume that all photons launched into the same input port have the same polarisation state. Eq.~\ref{MportRepDiffPol} then simplifies to 
\begin{equation}
    |\Psi_{\Vec{n}}(\Vec{\theta},\Vec{\phi})\rangle = 
    \prod_{m=1}^{M}\frac{a_m^{\dag}{(\theta_m,\phi_m)}^{n_m}}{ \sqrt{n_m!}}|0 \rangle.
\end{equation}

To represent the polarisation state explicitly in terms of the \((\theta_m,\phi_m)\) we write the state of the input or output in the \(2M\) mode occupation number basis. In this representation, each of the \(M\) ports has two associated modes representing the horizontally and vertically polarised components of the electric field. By applying the creation operator replacement rule \(a_m^{\dag}(\theta_m,\phi_m) \rightarrow \cos{\theta_m}a_{m,H}^{\dag} +\sin{\theta_m}e^{i\phi_m}a_{m,V}^{\dag}\), the state of the field can be written as
\begin{equation}
    \label{2Mrep}
    |\Psi_{\Vec{n}}(\Vec{\theta},\Vec{\phi})\rangle = 
    \prod_{m=1}^{M}\frac{{(\cos{\theta_m}a_{m,H}^{\dag} +\sin{\theta_m}e^{i\phi_m}a_{m,V}^{\dag})}^{n_m} }{ \sqrt{n_m!}}|0 \rangle,
\end{equation}
where \(a_{m,H/V}^{\dag}\) denotes a creation operator acting on the horizontally/vertically polarised component of port \(m\). The coefficients of these operators may be converted to Poincar\'e sphere coordinates.

\section{Model of polarising network}
Quantum states of light may be propagated through a linear photonic network (LPN) by applying the method developed by Scheel~\cite{Scheel2004}, which takes as input the unitary scattering matrix \(\Lambda\) whose elements are the coupling amplitudes from each input mode to each output mode of the {LPN}.

Without loss of generality, we construct a LPN using the design of Reck et al.\ as shown in Supplementary Figure~\ref{Mport}. It is composed of a triangular arrangement of \(Z = M(M-1)/2\) beamsplitters and \(M(M+1)/2\) phase-shifters. Such a network, if lossless, can physically realise any discrete \(M \times M\) unitary transform~\cite{Reck1994}.

\begin{figure}[!htbp]
    \begin{center}
        \includegraphics[width = 8cm]{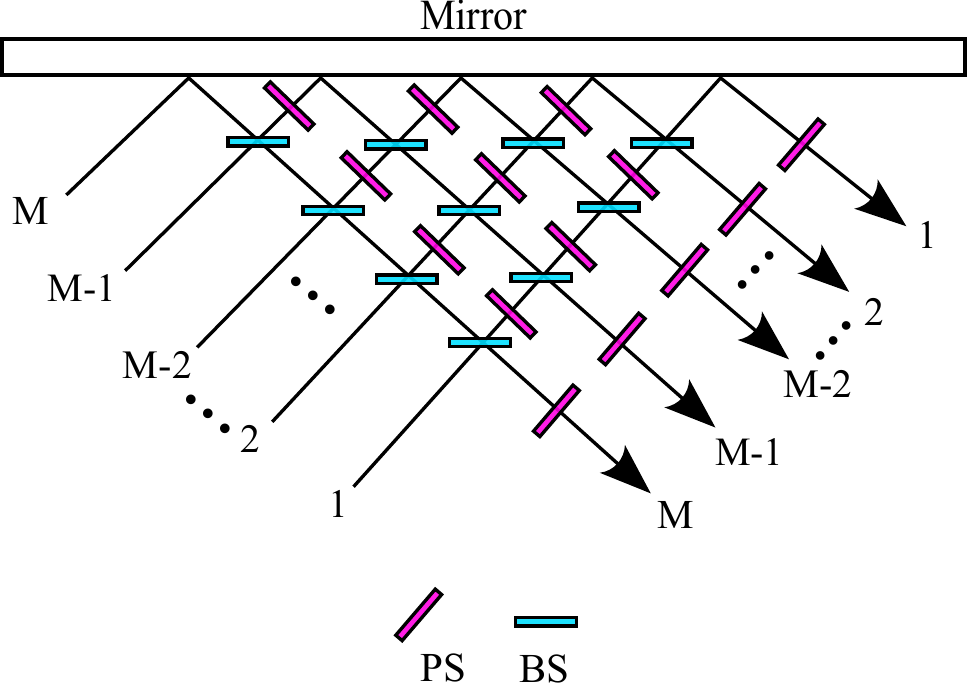}
    \end{center}
    \caption{M-port LPN using triangular arrangement of beamsplitters (BS) and phase shifters (PS).}\label{Mport}
\end{figure}

The \(\Lambda\) matrix is built from the ordered product of beamsplitter matrices which represent the coupling between two spatial modes at the crossing points of the interferometer. A single lossless beamsplitter followed by a phase shifter may be represented by a unitary matrix with two free parameters as follows:
\begin{equation}
    \begin{aligned}
    C(\alpha,\beta) = \begin{pmatrix}
            \cos{\alpha} e^{i\beta} & \sin{\alpha} \\
            -\sin{\alpha} e^{i\beta} & \cos{\alpha} \\
        \end{pmatrix},
    \end{aligned}
\end{equation}
where \(\cos\alpha = t\) and \(\sin\alpha = r\) are the transmission and reflectance coefficients of the beamsplitter and \(\beta\) is the phase accrued after propagation through the phase-shifter. To apply this rotation in an M-dimensional space, we use the method of Givens rotations and embed the two-dimensional rotation in an \(M \times M\) identity matrix so that it operates only in the appropriate two-dimensional subspace. The Givens rotation corresponding to a beamsplitter/phase-shifter combination operating on modes \(i\) and \(j\) may be written as  
\begin{align}
    \lambda^{ij}(\alpha,\beta) &= g^{ij}(C) \\
    &= \begin{pmatrix}
        1      &  & \cdots    & & 0 \\
         & \lambda_{i,i} = C_{11} & & \lambda_{i,j} = C_{12} &  \\
        \vdots &                       & 1 &                   & \vdots \\
         & \lambda_{j,i} = C_{21} &  & \lambda_{j,j} = C_{22} &  \\
        0      &  &\cdots  & & 1 \\
    \end{pmatrix},
\end{align}
where we use \(g^{ij}(C)\) to denote the Givens rotation generated from the matrix \(C\). Next, we define a diagonal matrix, \(\Psi\) whose entries contain the final output phase shifts for each mode:
\begin{equation}
    \Psi = \begin{pmatrix}
        e^{i\psi_1} &    0   & \cdots  & 0 \\
         0     & e^{i\psi_2} &  \cdots & 0 \\
        \vdots &             & \ddots &\vdots \\
         0     & 0           & \cdots  & e^{i\psi_M}\\
    \end{pmatrix}.
\end{equation}

The full unitary transformation in the space of modes, \(\Lambda\) is built by an ordered product of \(Z\) two-mode coupling matrices followed by \(\Psi\) as follows:
\begin{equation}
   \label{ScatMatFact}
   \begin{aligned}
       \Lambda = \Psi \left[\lambda^{M-1,M} \right]\left[\lambda^{M-2,M-1}\lambda^{M-2,M} \right]\ldots\\
       \ldots\left[\lambda^{1,2}\lambda^{1,3}\ldots\lambda^{1,M-1}\lambda^{1,M} \right],
   \end{aligned}
\end{equation}
where ordering is determined by the labelling of modes in Figure~\ref{Mport}.

To model a polarising LPN we construct \(\Lambda\) in the \(2M\) dimensional space formed from the tensor product of the state space of \(M\) spatial modes with the 2-dimensional state space of polarisation. We follow the same procedure as before, however, the primitive elements are now imperfect polarising beamsplitters, birefringent crystals and phase shifters operating in the tensor product space. The \(4 \times 4\) matrices defining these transformations are:
\begin{equation}
    \begin{aligned}
        \text{IPBS} &= \ \begin{blockarray}{ccccc}
            m_H & p_H & m_V & p_V\\
            \begin{block}{(cccc)l}
            \cos{\alpha_h} & \sin{\alpha_h} & 0 & 0 & \ \ m_H\\
            -\sin{\alpha_h} & \cos{\alpha_h} & 0 & 0 & \ \ p_H\\
            0 & 0 & \cos{\alpha_v} & \sin{\alpha_v}& \ \ m_V\\
            0 & 0 & -\sin{\alpha_v} & \cos{\alpha_v}& \ \ p_V\\
            \end{block}
        \end{blockarray},
    \end{aligned}
    \label{IPBS}
\end{equation}
where \(\alpha_{h/v}\) is the beamsplitter reflectance parameter for H/V polarisation and the rows and columns are labelled to make explicit the coupling between the H and V subspaces. We can see that the transformation in Eq.~\ref{IPBS} couples different polarisations to different spatial modes (ports) but does not mix the amplitudes of the internal state components. To accomplish this we need birefringent materials. Following the same mode ordering as before, the transformation induced by placing two birefringent plates at the outputs of a beamsplitter may be represented as 
\[
    \resizebox{1.0\hsize}{!}{%
        $
        R = \begin{pmatrix}
             e^{-i\eta_1/2} (\cos{\theta_1}^2 + e^{{-i\eta_1}}\sin{\theta_1}^2) & 0 & e^{-i\eta_1/2} (1-e^{i\eta_1} e^{-i\phi_1}\cos{\theta_1}\sin{\theta_1} & 0\\
            0 & e^{{-i\eta_2/2}} (\cos{\theta_2}^2 + e^{{-i\eta_2}}\sin{\theta_2}^2) & 0 & e^{-i\eta_2/2} (1-e^{i\eta_2}) e^{-i\phi_2}\cos{\theta_2}\sin{\theta_2}\\
            e^{-i\eta_1/2} (1-e^{i\eta_1}) e^{i\phi_1}\cos{\theta_1}\sin{\theta_1} & 0 & e^{{-i\eta_1/}2} (\cos{\theta_1}^2 + e^{{-i\eta_1}}\sin{\theta_1}^2) & 0\\
            0 & e^{-i\eta_2/2} (1-e^{i\eta_2}) e^{i\phi_2}\cos{\theta_2}\sin{\theta_2} & 0 & e^{{-i\eta_2/2}} (\cos{\theta_2}^2 + e^{{-i\eta_2}}\sin{\theta_2}^2)\\
        \end{pmatrix},%
        $
    }%
\]
where \(\eta_i\) is the linear retardance, \(\phi_i\) is the circularity and \(\theta_i\) is the waveplate angle. The index, \(i\) denotes the output port of the beamsplitter. The rows and columns denote the same coupling of subspaces as in Eq.~\ref{IPBS}. We can then create the Givens rotations corresponding to these \(4\times4\) matrices by including four superscripts instead of two signifying that \(16\) terms will be replaced in the \(2M\times2M\) dimensional identity matrix like
\begin{equation}
    \lambda^{mp} = g^{m_H,m_V,p_H,p_V}(C),    
\end{equation}
where \(C\) is now the \(4\times4\) coupling matrix between spatial modes \(m\) and \(p\). The aggregate transformation of an IPBS followed by two birefringent plates followed by a single phase shifter is then written
\begin{align*}
    \lambda^{mp} = g^{m_H,m_V,p_H,p_V}(PS)*g^{m_H,m_V,p_H,p_V}(R)*\cdots\\
    \cdots*g^{m_H,m_V,p_H,p_V}(\text{IPBS}).
\end{align*}

We can then build the full \(2M\times2M\) network matrix by multiplying the \(Z\) \(\lambda^{mp}\) matrices according to the crossing point order specified in Eq.~\ref{ScatMatFact}.
While this presents a general model of a polarising network, in the results presented we consider non-polarising, variable beamsplitters. 

\begin{figure}[b!]
    \begin{center}
        \includegraphics[width=0.6\columnwidth]{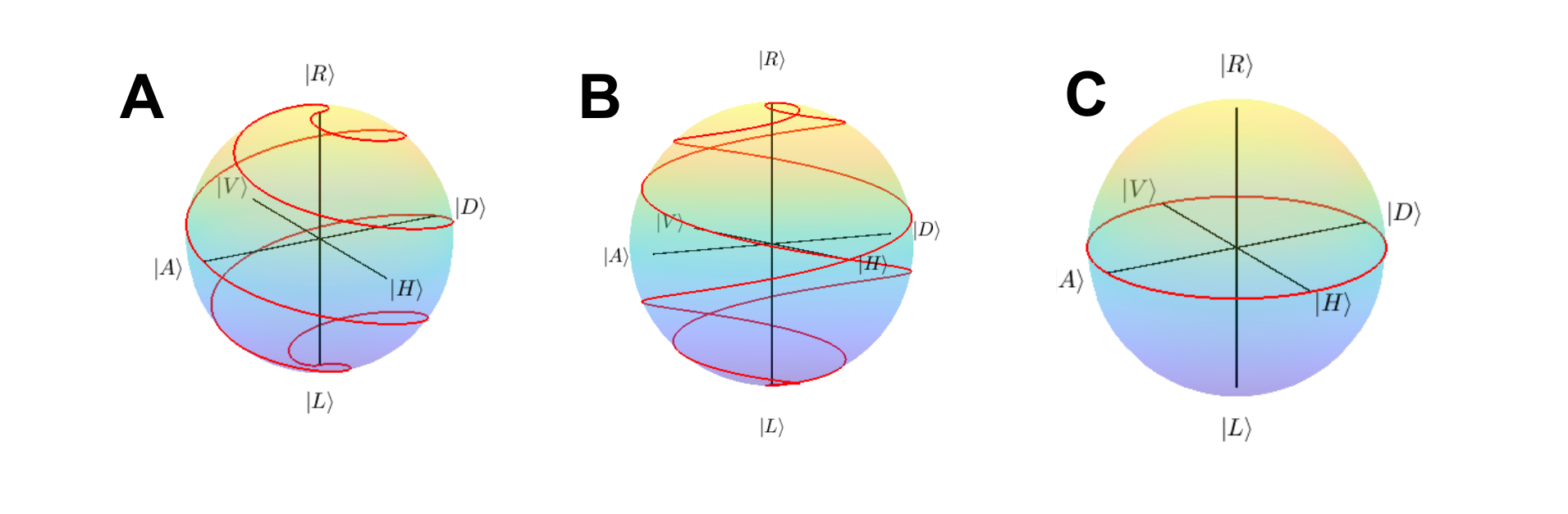}
        \caption{Poincar\'e sphere representations of a single port's polarisation trajectory over the data domain for various encoding schemes. Encodings considered here are: A. interleaved spiral (\(\xi = 1, \gamma = 1, \nu = 3\)) B. overlapping spiral (\(\xi = 1, \gamma = 0, \nu = 3\))  C. linear (\(\xi = 0, \gamma = 0, \nu = .5\)).}\label{diffencodings}
    \end{center}
\end{figure}

\section{Encoding schemes}
The first step of operating a machine learning model is to encode input data \(x\) in such a way that the device can process it. Note that while we consider scalar \(x\), the scheme is generalisable to multivariate inputs through encoding schemes which distribute input data elements across input modes. Encoding data into the LPN is achieved by feeding a fixed, horizontally polarised input state into an encoding layer composed of a quarter- and half-waveplate placed before each input port. To build the scattering matrix of this layer we multiply \(M\) Givens rotations generated by the Jones matrices of the two encoding waveplates at each port, denoted \(J( \theta_{Q,m}({\bf x}), \theta_{H,m}({\bf x}))\). These rotations act only on the \(2 \times 2\) subspace spanned by the polarisation modes \((m_H,m_V)\) at each port. Thus, the resulting matrices are block diagonal and commute.
\begin{equation}
    E({\bf x}) = \prod_m^M g^{m_H,m_V}( J(\theta_{Q,m}({\bf x}), \theta_{H,m}({\bf x})))
\end{equation}

An encoding scheme then is realised by a set of \(2M\) waveplate angles \((\theta_{Q,m}({\bf x}), \theta_{H,m}({\bf x}))\) which are functions of the data. These functions create states which follow closed trajectories on the Poincar\'e sphere at the corresponding port and the choice of an encoding scheme determines the character of the output functions of the device~\cite{Schuld2021,Gan2022,GilVidal2020}. A complete characterisation and optimisation over possible encodings is beyond the scope of this work, however, we propose a few natural options here and provide some analysis in order to clarify some details of the main text.

\begin{figure*}[b!]
\begin{center}
\includegraphics[width=.8\columnwidth]{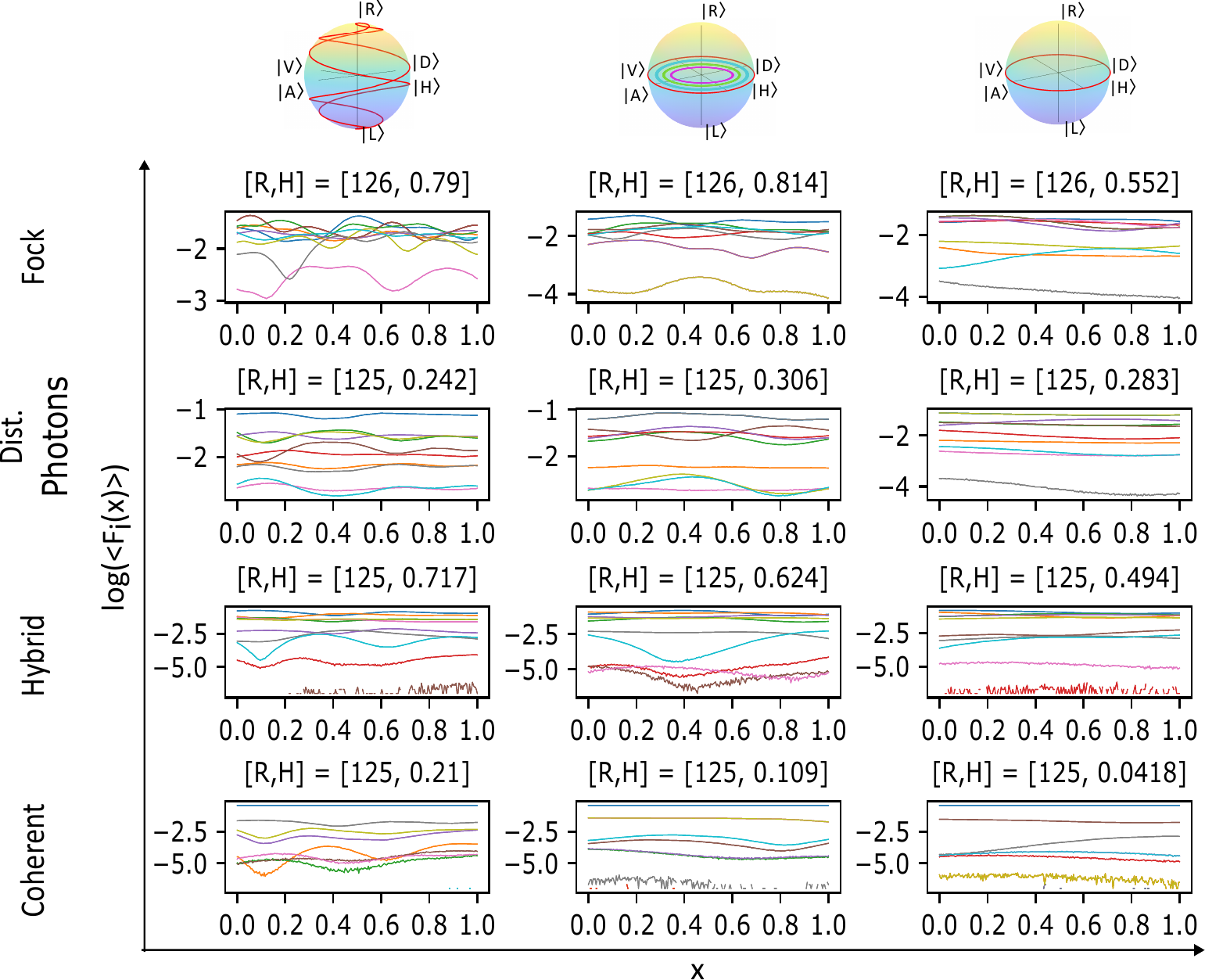}
\caption{Ten most probable output functions for three different states and encodings post-selected for detection events of \(\leq 4\) photons with \(\eta = .9\) and \(N_{samp} = 10^7\). The states used are Fock: \(|1,1,1,1,0 \rangle\), Distinguishable four photon state: \(|1_d,1_d,1_d,1_d,0 \rangle\), Hybrid: \(|\alpha = .5 + 1,\alpha = .5 + 1,0,0,0 \rangle\) and Coherent: \(|\alpha = 1.5,\alpha = 1.5,0,0,0 \rangle\). All configurations are subject to PNR detection Encodings are Uniform Linear (\(\xi = 0, \gamma = 0, \nu_m = .5\)), Multi-Linear (\(\xi = 0, \gamma = 0, \nu_m = m\)) and Spiral (\(\xi = 1, \gamma = 0, \nu_m = 4m\)).}\label{FxnDistributions}
\end{center}
\end{figure*}

All encodings used in this work are classes of closed spirals as shown in Supplementary Figure~\ref{diffencodings}. As such we have unified them with a single parametrisation, periodic on the interval \([-1,1]\) with degrees of freedom for each port \(m\)
\begin{align*}
    x' &= x + \rho_m \\ 
    \theta_{Q,m}(x) &= \xi_m(1 + 2x' - 4x'\gamma_m\mathcal{H}(x'))\frac{\pi}{4}\\
    \theta_{H,m}(x) &= (\nu_{m}x' + \xi_m(\gamma_m\mathcal{H}(x')) + \cdots\\
    & \cdots + 2\nu_{m}x'\gamma_m\mathcal{H}(x') - 2\nu_{m}x')\frac{\pi}{4}
\end{align*}
where \(\mathcal{H}\) is the Heaviside step function, \(\nu_m\) determines the number of azimuthal orbits the spiral will complete and \(\xi_m, \gamma_m \in \{0,1\}\) determine if the trajectory leaves the equator and if the spiral reverses direction at the poles to interleave, respectively. The variable \(\rho_m\) is a phase offset in the interval \([0,2]\) and sets the starting point of each encoding trajectory on the Poincar\'e sphere. This creates output functions with more phase diversity and increases the performance of fitting.

\section{Examples of interpolation of common target functions}
\begin{figure}[htp!]
\begin{center}
\includegraphics[width=1.0\columnwidth]{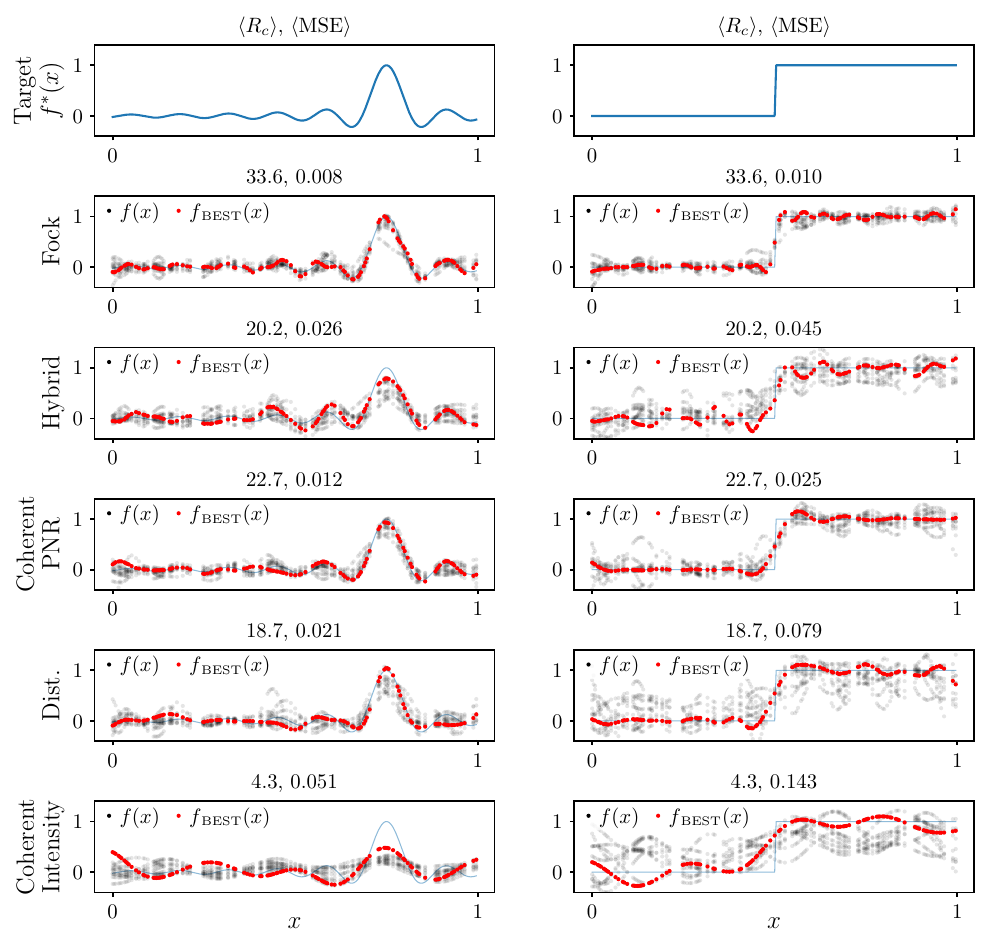}
\caption{{\bf{Function interpolation}}, comparing the approximation \(f(x)\) to the target function \(f^*(x)\) for ten different random reservoirs. Each subplot shows the result of the best fit in red while the shadows display results from other reservoir realisations. All trials use spiral encoding with \(l_m = 4m\) and post-selection on \(\leq 4\) photon events with \(\eta = .9\) and \(N_{samp} = 10^7\). Performance metrics [\(\langle R_c\rangle \),\(\langle MSE\rangle \)] are listed in the title for each task/ state combination. The states used are {\bf Fock:} \(|\vec{n} \rangle=|1,1,1,1,0 \rangle\) with PNR detection, {\bf Hybrid:} \(|\psi_{\vec{\alpha},\vec{n}} \rangle=|0.5 + 1,0.5 + 1,0,0,0 \rangle\) with PNR detection, {\bf Coherent PNR:} \(|\vec{\alpha} \rangle = |0.5,0.5,0,0,0\rangle\) with PNR detection, {\bf Distinguishable:} \(|\vec{n}_d \rangle=|1_d,1_d,1_d,1_d,0 \rangle\) with PNR detection, {\bf Coherent Intensity:} Fully classical operation using \(|\vec{\alpha} \rangle = |0.5,0.5,0.5,0.5,0\rangle\) with intensity detection. Each coherent state subject to PNR detection is approximated as a superposition of Fock states up to the sixth order. Each trial uses a 50/50 test/train data split.}\label{BenchmarkInterp}
\end{center}
\end{figure}

\section{Image classification details}\label{sec:image_classification}

We perform image classification on the digit MNIST and a subset of six MedMNIST datasets. 
Each task features between 2 and 11 classes. 
In order to efficiently train the reservoir, we limit ourselves to at most 10,000 training samples and evaluate on 1000 unseen test samples for each task.
All datasets feature images of size \(28\times28\), with the `BloodMNIST', `PathMNIST' and `DermaMNIST' datasets also having 3 colour channels.
In each case, to efficiently encode a given image in the QRC, we perform principal component analysis over the training dataset and select the first 20 principal components. This remains a fully linear dimensionality reduction of the data, and therefore does not provide any {performance improvement} over encoding the raw pixel data. While it may limit the overall performance on a given task, here we are interested in the relative performance of PNR detection and different input states, not overall performance on a given task.

\begin{figure*}[tb]
	\begin{center}
			\includegraphics[width=1.0\columnwidth]{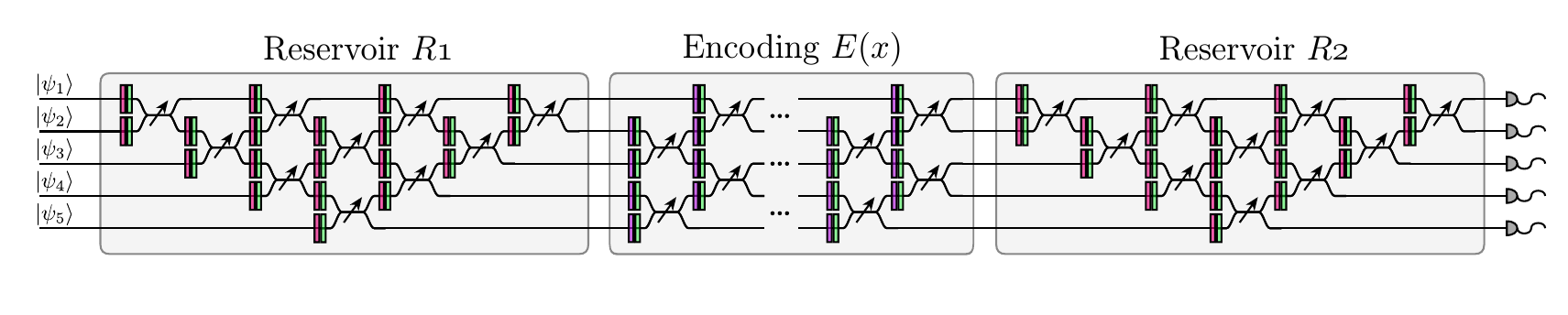}
			\caption{
					{\bf{Image classification reservoir scheme.}}
                    The first 20 principal components (PCs) of an input data point are encoded in the first 20 variable beamsplitters in the encoding block \(E(x)\), where the PCs are normalised to the range \([\pi/4, 3\pi/4]\). All waveplates and phase shifts remain random.
			}\label{fig:image_mesh}
	\end{center}
\end{figure*}

In order to encode the necessary number of values (20 principal components), we modify the reservoir scheme slightly, by adding a secondary random unitary before the encoding layer. Additionally, we change from a polarisation encoding, to one which uses the same waveplate and variable beamsplitter geometry as the random reservoirs. Here, the waveplates and phase shifts are left random, but the beamsplitter angles are varied according to the principal components, which are normalised to the range \([\pi/4, 3\pi/4]\). The modified reservoir is shown in Figure~\ref{fig:image_mesh}.

While this may seem a large departure from the scheme depicted in Figure 1, the addition of the first random unitary only serves to distribute our limited number of photons across all modes. 
This could be accomplished, in photonic integrated circuits, or in a fibre based approach for instance, by injecting photons into an initial multimode fibre \(R_1\) before being encoded by a spatial light modulator \(E(x)\) and injected into a second fibre \(R_2\). Thus, the additional scrambling network is a trivial addition which largely preserves our architecture in practice. 
While the precise polarisation encoding used in function fitting allowed us to analyse the reservoir through the frequency components we were able to generate, it is not clear that a broad spectrum is necessary in classification tasks. The use of variable beamsplitters to encode information therefore serves to demonstrate the flexibility of our system to a wide range of substrates and encoding schemes.

\begin{figure*}[!b]
	\begin{center}
			\includegraphics[width=1.0\columnwidth, trim={0 0 0 6.35cm}, clip]{image_classification.pdf}
			\caption{
					{\bf{Image classification Matthews correlation coefficients (MCC).}} The majority classifier, which always predicts the dominant class from the training data, always gives MCC of zero, and a perfect classifier would give MCC of one.
			}\label{fig:image_mcc}
	\end{center}
\end{figure*}

Once the outputs \(\langle \hat{F}(x_i) \rangle\) have been generated from the reservoir, classification is performed by training the output weights to predict probabilities over the class labels, where final classification is done via maximum likelihood. The training process is identical to that shown in Figure 1.

One issue with evaluating performance in classification tasks is that overall accuracy is not a good metric if the dataset is imbalanced (i.e.\ classes are not uniformly sampled). Instead, we may use the Matthews correlation coefficients (MCC), shown in  Figure~\ref{fig:image_mcc}, which allows us to fairly compare the performance on unbalanced datasets. 

Low MCC in OCT, Breast and Derma datasets for all classifiers indicates that these tasks are particularly difficult due to class imbalance. In the case of Breast, we see that the Fock and coherent PNR classifiers overfit to the training data and perform poorly on test sets.
It is expected that, as with interpolation tasks, tailoring the encoding scheme to the task at hand will improve performance in these cases. 
Regardless of absolute performance, the hierarchy of PNR detection remains consistent across the tasks.

\section{Simulation of input states}
To simulate and compare input Fock, hybrid and coherent states subject to PNR detection we can express them in a unified form as 
\begin{equation}
  |\psi_{\vec{\alpha},\vec{n}} \rangle = \prod_{m=1}^M \frac{{(\hat{a}_m^{\dag})}^{n_m}D_m(\alpha_m)}{\sqrt{L_{n_m}(-{|\alpha|}^2)n_m!}}|0 \rangle
  \label{sup:hybridstate}
\end{equation}

Setting the \(\alpha_m = 0\) simplifies Eq~\ref{sup:hybridstate} to the standard form of a Fock state while setting the \(n_m = 0\) simplifies it to a multimode coherent state. To calculate the amplitudes of different PNR detection results we expand the \(D_m(\alpha_m)\) into the standard Poissonian-weighted power series of creation operators truncated at the order where the coefficient falls below one percent of the peak value. The result is a superposition over Fock states which we propagate through the network individually and recombine into a coherent sum at the output. In the case of polarised states, we apply the operator replacement rule from Section~\ref{sec:fock_states} and proceed in the same way. Since we do not consider polarisation in the detection phase, a partial trace is performed over the polarisation degree of freedom after propagating the full state through the network.

\section{Comparing states}

There are few degrees of freedom in a pure Fock state, however hybrid and coherent states are characterised by a set of continuous amplitudes \(\alpha_m\). Fair comparison of the performance of these three classes of states on a task presents a challenge as they have inherently different character. For instance, if we post select on events which contain four or fewer photons we can make a choice of amplitudes that maximises the probability mass in this subspace. However, a given target function may contain a Fourier spectrum which overlaps more with frequencies generated by three-photon events. This makes the choice of optimal \(\alpha_m\) task-specific rather than just depending on the post-selected subspace. In addition, detector losses couple higher number states to lower number states, which shifts the probability mass and must be accounted for. We can then treat the \(\alpha_m\) as hyperparameters which may be chosen based on the task at hand. In this work, we have chosen to keep the problem as general as possible and set the \(\alpha_m\) to maximise probability mass in the post-selected subspace rather than go through this process of specialisation.

\section{Detector model}\label{sec:detector}
To simulate output measurements under real-world conditions, we need to formulate a model for noise. Assuming perfect photon sources, noise in a LPN can arise in two ways: losses in the network and imperfections in the detectors. Although network losses can impact the output statistics it is common to neglect them for proof-of-principle simulations. This can be justified on the experimental side as LPNs can be manufactured with tolerances such that losses are negligible for modestly sized networks and also for practical reasons as simulation of inhomogeneous loss quickly becomes intractable. In this work we focus on the effect of detector imperfections including dark counts and detector losses. We consider the case of one PNR detector at each output port and assume independent noise processes for each. To justify neglecting dark noise in our model we make a conservative estimate of dark noise for a modern single photon detector as 100 cps. Assuming a coincidence window of 10 ns this yields an average of \( 10^{-6}\) dark counts per detector over this time period. Although modern detectors can exhibit negligible levels of dark noise even in the case of singles detection, the probability of a spurious detection event rapidly tends towards zero as we consider two-photon events and higher. Detector losses, however, can have a significant impact on the Photon-QuaRC output, even with high quantum efficiencies. We model the probability of recording \(n'\) counts given \(n\) input photons as a binomial process
\begin{equation}
    P(n' \;|\; n ) = \binom{n}{n - n'} {(1-\eta)}^{n-n'}\eta^{n'}
\end{equation}
where \(\eta\) is the detector quantum efficiency. This process maps an output Fock state \(|\vec{n}\rangle\) to a probability distribution over a subspace of Fock states \(|\vec{n}'\rangle\) with probabilities    
\begin{equation}
    P(\vec{n}'| \vec{n}) = \prod_m P(n'_m \;|\; n_m ).
\end{equation}

Once we propagate an input state through the network for the ideal case we then apply the binomial loss model to redistribute the probabilities over the Fock space with photon number less than or equal to \(N\).

\section{Post-selection as a hyperparameter}\label{sec:postsel}
To clarify the significance of post-selection in Photon-QuaRC let us consider an example. Suppose we choose a coherent source with amplitude \(\alpha\) and some encoding to fit a target function. From main text Eq. 4, the frequencies of the output functions will vary with the number of photons in their corresponding detection events. The Fourier spectrum of the target function might overlap only with that of the subset of \(\langle {\hat{F}}(x) \rangle\) corresponding to two and three photon Fock states. However, due to the nature of coherent states, a given detection event may fall well outside this range, effectively contributing noise to the fit. Thus, to reduce the burden of sampling and matrix inversion, and improve fit quality, we would only select detection events corresponding to the \(N = \left[2,3\right]\) subspace and reject all others. Misdetection of higher photon number states due to detector imperfections can also reduce the quality of predictions by corrupting the frequency content of our measurements. These considerations determine our requirements for detector noise and quantum efficiency while knowledge of the frequency range of our target function informs our choice of encoding, source and post-selection.

In all experiments shown, and for all states, we use the same post-selection---detection events with four or fewer total photons. This choice is made to ensure a fair comparison between the different states and encoding schemes, with four photons being the maximum in the Fock and distinguishable states tested. In a real-world scenario, the choice of post-selection would be task-dependent and would be made based on the frequency content of the target function.